\documentclass[aps,pre,twocolumn,groupedaddress,showkeys]{revtex4-1}

\usepackage{graphicx}
\usepackage{mathtools}%\usepackage{amsmath}

\begin{document}

% \preprint{}

\title{On Causality in Dynamical Systems}


\author{Daniel Harnack}
\email[]{daniel@neuro.uni-bremen.de}
\homepage[]{http://neuro.uni-bremen.de}
%\thanks{}
% \altaffiliation{ Center for Cognitive Sciences}
\affiliation{University of Bremen, Institute for Theoretical Physics}
\affiliation{Center for Cognitive Science (ZKW)}
\author{Erik Laminski}
\email[]{e.laminski@uni-bremen.de}
%\homepage[]{http://neuro.uni-bremen.de}
%\thanks{}
% \altaffiliation{ Center for Cognitive Sciences}
\affiliation{University of Bremen, Institute for Theoretical Physics}
\affiliation{Center for Cognitive Science (ZKW)}
\author{Klaus Richard Pawelzik}
\email{pawelzik@neuro.uni-bremen.de}
\homepage[]{http://neuro.uni-bremen.de}
%\thanks{}
\affiliation{University of Bremen, Institute for Theoretical Physics}
\affiliation{Center for Cognitive Science (ZKW)}


\date{\today}

\begin{abstract}
Discovery of causal relations is fundamental for understanding the dynamics of complex systems. While causal interactions are well defined for acyclic systems that can be separated into causally effective subsystems, a mathematical definition of gradual causal interaction is still lacking for non-separable dynamical systems. The solution proposed here is analytically tractable for time discrete chaotic maps and is shown to fulfill basic requirements for causality measures. It implies a method for determination of directed effective influences using pairs of measurements from dynamical systems. Applications to time series from systems of coupled differential equations and linear stochastic systems demonstrate its general utility. 
\end{abstract}



\maketitle

\section*{Introduction}
The notion of causality has a long history ranging back to ancient philosophers including Aristotle \citep{aristotle}. In recent formalizations it refers to situations where states $x_1$ of one part of a system influence the states $x_2$ of some other part \citep{Ay_2008}. It is further assumed that some aspects of $x_1$ vary independently of $x_2$, and that the flow of information in the overall system is essentially unidirectional. This premise of acyclic interaction is at odds with complex dynamical systems studied in e.g. ecology, economy, climatology and neuroscience: generally, two system parts, e.g. two brain areas, will have bidirectional interaction and cyclic information flow. The classical notion of causality becomes problematic here since cause and effect are entangled.

This entanglement is reflected in Takens' theorem \citep{Takens_1981, Packard_1980}, which proves that in deterministic dynamical systems the overall state is reconstructible from any measured observable using time-delay coordinates. In other words, if $x_1$ and $x_2$ interact bidirectionally, each time series $x_1(t)$ and $x_2(t)$ contains the full information about the whole system made up of $x_1$ and $x_2$. That is, the system cannot be separated into subsystems and rather behaves as a whole. In consequence, the question for causal relations in such a system can not be answered by a classification of component systems into cause and effect, but rather asks for the directed effective influence between these component systems.

Here we present a mathematical definition of directed effective influence tailored to entangled dynamical systems which is based on topological considerations. As a key insight we discovered that local distortions in the mappings between reconstructions based on different component systems directly reflect the time dependent efficacy of causal links among these components. A causality index derived from this relation, which we term 'Topological Causality', is analytically accessible for simple systems and can be estimated in a model free, data driven manner for more complicated ones. We propose this measure as a suitable extension of the causality concept to non-separable dynamical systems.

\section*{Results} 
The concept of Topological Causality introduced here relies on Takens' theorem, which will be reviewed shortly with an example. Let variables $x_1$ and $x_2$ be governed by dynamical equations
\begin{align*}
\dot{x}_1 &= f_1(x_1,w_{12}x_2) \\
\dot{x}_2 &= f_2(x_2,w_{21}x_1) 
\end{align*}
The system generates trajectories $(x_1(t),x_2(t))$ over time $t$ which for dissipative systems lie on specifically shaped manifolds. Takens' theorem states that these manifolds are topologically equivalent to manifolds visited by $r^{x_1}(t) = (x_1(t),x_1(t+\tau),x_1(t+2\tau)\dots, x_1(t+(m-1)\tau))$ in a delay coordinate space if $w_{12} \neq 0$ and the embedding dimension $m$ is sufficient. The same holds for reconstructions $r^{x_2}$ based on $x_2$ if $w_{21} \neq 0$. 

Topological equivalence of manifolds means that homeomorphic, neighborhood preserving one-to-one mappings exist between these manifolds. If both $w_{12} \neq 0$ and $w_{21} \neq 0$, also homeomorphic one-to-one mappings between reconstructions $r^{x_1}$ based on $x_1$ and $r^{x_2}$ based on $x_2$ exist. These mappings between reconstructions, e.g. from $r^{x_1}$ to $r^{x_2}$ denoted by $M_{1 \to 2}$, are the main objects of study.

\begin{figure}
	\includegraphics{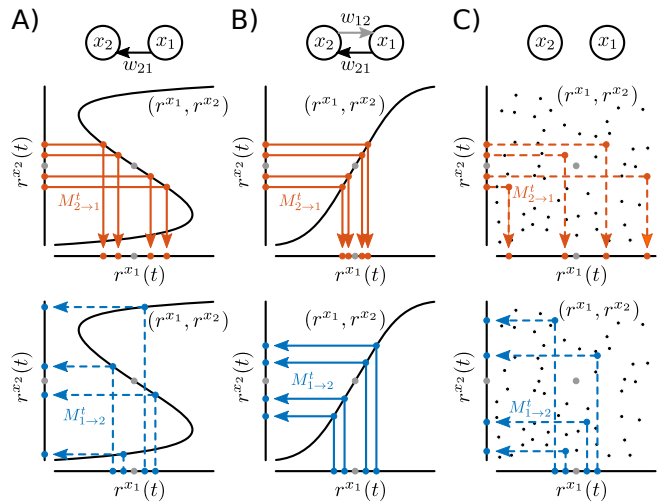}%
	\caption{\label{fig_idee} The relation of points $r^{x_1}$ and $r^{x_2}$ on multidimensional manifolds illustrated in 1-d. The joint manifold represented by $(r^{x_1},r^{x_2})$ can be interpreted as the function mediating the mappings $M$ between both spaces, and local linearizations $M^t$ of the mappings as the slope around a reference point. A) When only $w_{21} \neq 0$, a one-to-one mapping $M_{2 \to 1}$ from $r^{x_{2}}$ to $r^{x_{1}}$ exists, but not in the reverse direction: not for all states $r^{x_{1}}(t)$,  $r^{x_{2}}(t)$ is uniquely determined. Locally, $M^t_{1\to 2}$ can be attributed a diverging expansion property: close neighbors of a given point $r^{x_{1}}(t)$ map to distant parts of the joint density $(r^{x_1},r^{x_2})$ i.e. local expansion extends to macroscopic scales. The dashed lines visualize the non-uniqueness. B) Here, both couplings are non-zero, but $w_{21} > w_{12}$. Larger independence of $x_{1}$ implies a stronger expansion by $M_{1\to 2}$ than by $M_{2\to 1}$ at most reference points, which is indicated by the higher slope of $(r^{x_1},r^{x_2})$ when seen from $r^{x_1}$. C) If no coupling exists, expansion diverges in both directions.}
\end{figure}

To illustrate how properties of these mappings relate to directed effective influence, we first consider a case of unidirectional coupling from $x_1$ to $x_2$ ($w_{12} = 0$ and $w_{21} \neq 0$). Takens' theorem ensures that the overall state of the full system is contained in reconstructions $r^{x_{2}}$ based on $x_2$ alone. Moreover, a unique mapping $M_{2 \to 1}$ from reconstructions $r^{x_{2}}$ to $r^{x_{1}}$ exists. In the reverse direction, a unique mapping $M_{1 \to 2}$ does not exist, since $x_1$ has no information on $x_2$. This is schematically illustrated in Fig. \ref{fig_idee} A) by a joint manifold $(r^{x_{1}},r^{x_{2}})$ lying 'folded' over $r^{x_1}$ but uniquely over $r^{x_2}$. 

In practice, we will analyze properties of localized linearizations of these mappings around a reference point, denoted by $M^t$ (i.e. the Jacobian matrix). Given that $\{t^{x_{1}}_{1}, ..., t^{x_{1}}_{k}\}$ are the time indices of the nearest neighbors on $r^{x_1}$ to the reference point $r^{x_1}(t)$,  $M^{t}_{1\to 2}$ is approximated by the linear mapping which projects $\{r^{x_{1}}\left(t^{x_{1}}_{1}\right), ..., r^{x_{1}}\left(t^{x_{1}}_{k}\right)\}$ to $\{r^{x_{2}}\left(t^{x_{1}}_{1}\right), ..., r^{x_{2}}\left(t^{x_{1}}_{k}\right)\}$. Fig. \ref{fig_idee} A) illustrates the well defined mapping $M^{t}_{2\to 1}$, while in the inverse direction $M^{t}_{1\to 2}$ does not exist, at least not in the usual sense of uniqueness. Note here that somewhat counter-intuitively the influence from $x_1$ to $x_2$ is reflected in the 'backward' mapping $M_{2\to 1}$: the existence of a mapping $M_{2\to 1}$ implies the existence of coupling from $x_1$ to $x_2$.

We now further argue that not only the existence, but also the efficacy of directed influences is reflected in these mappings. More precisely, we postulate that the strength of the state dependent directed effective influence from $x_2(t)$ to $x_1(t)$ correlates with the degree of expansion of the mapping $M^{t}_{1 \to 2}$. The expansion $e^t$ of a mapping $M^t$ is determined by the $N$ singular values $\lambda_i$ of $M^t$ which are larger than one:
\begin{equation}
\label{eq:expand}
e^t = \prod_{i=1}^{N}\lambda_{i} \\
\end{equation} 
This entails that the more expanding $M^{t}_{1 \to 2}$ is, the bigger the distances within the corresponding set of points $\{r^{x_{2}}\left(t^{x_{1}}_{1}\right), ..., r^{x_{2}}\left(t^{x_{1}}_{k}\right)\}$ on $r^{x_{2}}$ will be in relation to the distances between $\{r^{x_{1}}\left(t^{x_{1}}_{1}\right), ..., r^{x_{1}}\left(t^{x_{1}}_{k}\right)\}$.

Staying in the previous example illustrated in Fig. \ref{fig_idee} A), one sees that the expansion of $M^t_{1\to 2}$ is quite large since the corresponding points lie scattered over the whole dynamical range of $r^{x_2}$. In the reverse direction the expansion will be smaller since the trajectory of $r^{x_2}$ contains information from and is thus constrained by $r^{x_1}$. As already noted above, in this limiting case of vanishing coupling ($w_{12} = 0$) a unique mapping $M^t_{1\to 2}$ does not exist in a strict mathematical sense. Still, this non-uniqueness can be identified with an infinite expansion property: In the limit of infinite observations, the distances of the nearest neighbors to the reference point $r^{x_1}(t)$ will approach zero whereas the distances of the corresponding points on $r^{x_2}$ will not decrease.

To further elaborate on the notion that weaker directed influence corresponds to stronger expansion consider the case that both couplings are nonzero, but $w_{21} > w_{12}$. Now state reconstructions $r^{x_{1}}$ and $r^{x_{2}}$ will both reveal the same global system state and are therefore topologically equivalent. However, the weaker coupling from $x_2$ to $x_1$ implies that the homeomorphic mapping $M^t_{1\to 2}$ will be more expanding than the mapping $M^t_{2\to 1}$ at most reference points, since movement of $x_2$ is less constrained by the influence of $x_1$ than vice versa. This can be visualized by the joint manifold $(r^{x_{1}},r^{x_{2}})$ lying uniquely over both reconstruction spaces, but more 'steeply' over $r^{x_{1}}$ (Fig. \ref{fig_idee} B)).
If the interaction strength $w_{12}$ is further decreased, one sees that while approaching the first case (Fig. \ref{fig_idee} A)) $M^t_{1\to 2}$ becomes more expansive and 'steeper' until it looses its uniqueness at $w_{12} = 0$ and the expansion diverges locally.

As a third concluding example consider the extreme case where $x_1$ and $x_2$ are completely decoupled, i.e. $w_{21} = w_{12} = 0$. Then both component systems will behave independently and the density of the resulting joint manifold factorizes. When observed from reference states $r^{x_{1}}(t)$ and $r^{x_{2}}(t)$, the mappings can be considered infinitely expanding, now in both directions, since for most reference points close neighbors correspond to distant points in the respective other space (Fig. \ref{fig_idee} C)).

Taken together, these topological considerations suggest that local expansions of the mappings between reconstruction manifolds of two observables might be utilized for a graded measure of directed causal influence between component systems represented by these observables.

\subsection*{An example guided definition of causality}

The putatively fundamental relation between effective influence and expansion can be analyzed in simple examples of coupled time discrete logistic maps described by 
\begin{align}
\label{eq:pawelzik}
x_{i}(t+1) &= \left(1-\sum_{j \in \{1, ...,n\}\backslash\{i\}}w_{ij} \right)f(x_{i}(t)) \\ \nonumber
&+ \sum_{j \in \{1, ...,n\}\backslash\{i\}}w_{ij}x_{j}(t) \nonumber
\end{align}
For the two-dimensional case ($n = 2$) with $f(x)= 4x(1-x)$, an embedding dimension $m=2$ is sufficient to reconstruct the full system state. Given the reconstruction states $r^{x_{1}}(t) = (x_{1}(t),x_{1}(t+1))$ and $r^{x_{2}}(t) = (x_{2}(t),x_{2}(t+1))$, the local mapping $M^t_{2\to 1}$, which projects a small area around the reference point $r^{x_{2}}(t)$ onto $r^{x_{1}}$, can be calculated. For small perturbations $\Delta_{x_{2}} = (\Delta x_{2}(t), \Delta x_{2}(t+1))$ around $r^{x_{2}}(t)$ one finds that 
\begin{equation*}
\Delta_{x_1} = \frac{1}{w_{21}} \left( \begin{matrix*}[c]
\tilde{f}_2 & & 1 \\
w_{12}w_{21}-\tilde{f}_1\tilde{f}_2 & & -\tilde{f}_1
\end{matrix*}\right) \Delta_{x_2}
\end{equation*}
with $\tilde{f}_2 = (w_{21}-1)f'(x_2(t))$ and $\tilde{f}_1 = (w_{12}-1)f'(x_1(t))$. This linearized perturbation matrix is equal to $M^t_{2\to 1}$ for $\Delta_{x_2} \to 0$ and both will therefore be used interchangeably in the following. Equivalently $M^t_{1\to 2}$ can be calculated. The expansions $e^t_{2\to 1}$ and $e^t_{1\to 2}$ are determined by Eq. \ref{eq:expand}. While the closed form solutions of $e^t_{2\to 1}$ and $e^t_{1\to 2}$ are quite unwieldy expressions, it can be seen that for small couplings $w_{21}$, $e^t_{2\to 1}$ is dominated by $1/{|w_{21}|}$ and vice versa. Thus, if $|w_{21}| > |w_{12}|$, the expansion $e^t_{2\to 1}$ of the mapping $M^t_{2\to 1}$ will be larger than $e^t_{1\to 2}$ (Fig. \ref{fig_idee} B)). This becomes most apparent for $w_{12} = 0$, in which case $e^t_{2\to 1}$ simplifies to 
\begin{equation}
\label{eq:stretching}
e^{t}_{2\to 1} = \frac{1}{|w_{21}|} \sqrt{ (w_{21}-1)^{2} f'^{2}(x_2(t)) + f'^{2}(x_1(t)) }
\end{equation}
This entails $\lim_{w_{21}\to 0} e_{2\to 1}^t = \infty$, confirming the intuition of infinite expansion for vanishing interaction (Figs. \ref{fig_idee} A), C)). For small $|w_{21}|$ the expansion $e^t$ of the mapping from $r^{x_2}$ to $r^{x_1}$ depends inversely on the coupling strength from $x_1$ to $x_2$ in this example (and vice versa). 

\begin{figure}
	\includegraphics{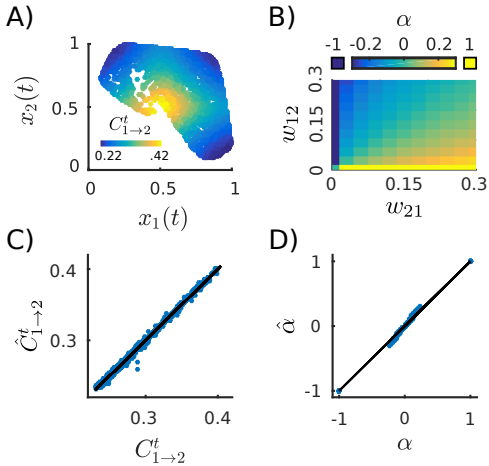}%
	\caption{\label{fig_linear_map} A) $C^{t}_{1\to 2}$ for the system described by Eqs. \ref{eq:pawelzik} $(n=2)$ with $w_{21}=0.3$ and $w_{12} = 0.23$. The map was iterated for 10.000 time steps and $C^{t}_{1\to 2}$ and $C^{t}_{2\to 1}$ calculated analytically at the points the system visited. B) The asymmetry index $\alpha$ for the same model as in A) with varying coupling weights. It accurately reflects the asymmetry in coupling weights such that $\alpha=0$ if $w_{21}=w_{12}$. C) The blue dots show a point to point comparison between analytical values for $C^{t}_{1\to 2}$ shown in A) and numerical estimates $\hat{C}^{t}_{1\to 2}$ and reveal a good agreement. A neighborhood size of $k = 10$ was used with a time series length of 10.000 data points and $m=2$. D) The theoretical values of $\alpha$ in B) are compared with the measured values $\hat{\alpha}$, and again a good agreement is found. Non-significant $\hat{C}$ were set to 0.}
\end{figure}

The expansion $e^t_{2\to 1}$ reflects increase of uncertainty induced by $M^t_{2\to 1}$.  From an information theoretical point of view the corresponding increase of entropy is bounded from above by $\log(e^t)$. Motivated by this interpretation we define the causality index as a ratio of uncertainties: 
\begin{equation*}
C^{t}_{1\to 2} = \frac{1}{1+\log\left(e_{2\to 1}^t\right)}
\end{equation*}
This definition, which we term Topological Causality (TC), satisfies the following intuitions about causality: First, TC from component system 1 to 2 vanishes if no causal link exists ($w_{21} = 0$). Secondly, for small couplings it is a monotonous function of the coupling weight $w_{21}$, at least in this simple example. However, there is a important distinction between TC and coupling weight: As defined here, $C^{t}$ depends on the coupling weights as well as on the current state of the system and the internal dynamics of each component (dependency on $x_1(t)$, $x_2(t)$ and $f()$ in Eq. (\ref{eq:stretching})). That is, coupling weights are static parameters that become effective in the context of the specific system. Fig. \ref{fig_linear_map} A) shows the state dependency of $C^{t}_{1\to 2}$ for a system described by Eqs. (\ref{eq:pawelzik}) $(n=2)$. 

Although $C^{t}_{1\to 2}$ and $C^{t}_{2\to 1}$ generally depend on the current state at time $t$, one might be interested in a global measure of causality reflecting the mean directed influence. For this the local expansion at every available state on the reconstructed manifolds can be averaged to yield 
\begin{equation*}
C_{1\to 2} := \frac{1}{T}\sum_{t=1}^{T}C^t_{1\to 2}
\end{equation*}
Furthermore, to address the asymmetry of causal influences between components 1 and 2, we define the local asymmetry index $\alpha^t$ as
\begin{equation*}
\alpha^t = \frac{C^{t}_{1\to 2} - C^{t}_{2\to 1}}{C^{t}_{1\to 2} + C^{t}_{2\to 1}}
\end{equation*}
and equivalently for the state-averaged values of $C$
\begin{equation*}
\alpha = \frac{C_{1\to 2} - C_{2\to 1}}{C_{1\to 2} + C_{2\to 1}}
\end{equation*}
Fig. \ref{fig_linear_map} B) displays $\alpha$ for a range of coupling parameters in a model described by Eqs. (\ref{eq:pawelzik}) $(n=2)$, showing that in this example the dominant mean influence is exerted along the stronger coupling weight. This is to be expected when both component systems are governed by the same model equations.

\subsection*{Estimating Topological Causality} 
In cases where the dynamical system model does not allow for an analytical linearization of the mappings between the reconstructed spaces, or the model itself is not known, the local mappings and hence their expansion can still be estimated in a purely data-driven manner. To estimate e.g. $M^t_{1\to 2}$ one finds the time indices $\{t^{x_{1}}_{1}, ..., t^{x_{1}}_{k}\}$ of the $k$ nearest neighbors on $r^{x_1}$ around a reference point $r^{x_1}(t)$. The projection from $\{r^{x_1}(t^{x_{1}}_{1}), ..., r^{x_1}(t^{x_{1}}_{k})\}$ to $\{r^{x_2}(t^{x_{1}}_{1}), ..., r^{x_2}(t^{x_{1}}_{k})\}$ is then approximately mediated by $M^t_{1\to 2}$ if the neighborhood size is sufficiently small. The approximation becomes exact in the limit of infinite observations. It is then straightforward to estimate $M^t_{1\to 2}$ by solving a simple optimization problem, e.g. by multivariate linear regression which is well documented in the literature (e.g. \citep{Anderson_1984}). Estimated values are denoted by a ' $\hat{}$ ' henceforth. Fig. \ref{fig_linear_map} C) and D) demonstrate that  $\hat{C}^{t}$, $\hat{C}$ and $\hat{\alpha}$ obtained in this manner are close to the theoretical values for a system given by Eq. (\ref{eq:pawelzik}) $(n=2)$. The same sets of points can also be used to fit the inverse matrices of $M^{t}$. Then the singular values smaller than one are taken into account for estimation of the expansion. We found that this latter procedure often yields more reliable results and therefore used it for the remaining numerics in this paper. Significance and chance level of the estimated $C^t$ and $C$-values were obtained by fitting matrices with the same sets of points where the time indices in the projection space were randomly permuted. Estimations were performed on time series with removed mean and normalized standard deviation.

\begin{figure}
	\includegraphics[]{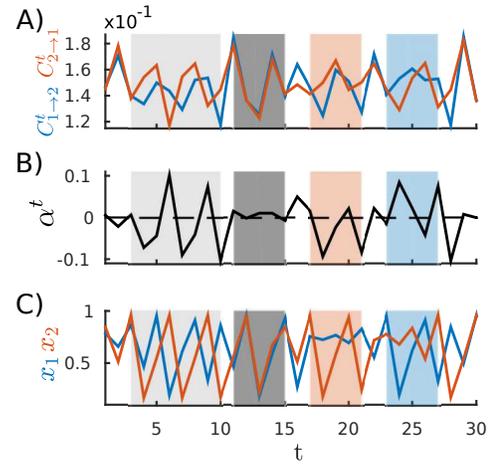}%
	\caption{\label{fig_sugihara} A) Time dependence of $C^{t}_{1\to 2}$ and $C^{t}_{2\to 1}$ in the system governed by Eqs. (\ref{eq:time_dependent}) with $w_{21} = w_{12} = 0.024$. B) The asymmetry index $\alpha^t$ shows a time dependency as well. Several regions are highlighted: A region in which the mean over time of $\alpha^{t}$ is close to 0 and the variance is high (light gray), one with similar mean but low variance (dark gray), one in which the mean is $<0$ (orange), indicating stronger influence $2\to 1$ than $1\to 2$, and one in which the mean is $>0$ (blue), indicating stronger influence $1\to 2$ than $2\to 1$. The same regions are marked accordingly in C) and A). C) Original time series of $x_1$ and $x_2$. The different regimes of causal asymmetry, marked by the shadings, give rise to different dynamical motifs. If $\alpha^{t}$ varies strongly around 0 (light gray), $x_1$ and $x_2$ desynchronize. If $\alpha^{t}$ is close to 0 for subsequent time points (dark gray), $x_1$ and $x_2$ synchronize. When the causal influence from $2 \to 1$ is dominant (orange), $x_2$ is more independent while movement of $x_1$ is constrained. The reverse is seen when the influence $1\to 2$ is dominant (blue).}
\end{figure}

\subsection*{Time dependent causal asymmetry}
Since the influence of a component system onto another may be state dependent, as evident from Eq. (\ref{eq:stretching}), so can the asymmetry index $\alpha^t$. This phenomenon can be investigated in coupled time-discrete maps that are non-linearly coupled. As an example consider the system given by
\begin{align}
\label{eq:time_dependent}
x_1(t+1) &= x_1(t)[3.8(1 - x_1(t)) - w_{12}x_2(t)] \\ \nonumber
x_2(t+1) &= x_2(t)[3.8(1 - x_2(t)) - w_{21}x_1(t)] \\ \nonumber
\end{align}
which may serve as a model of ecological systems \citep{Sugihara_2012}. Also for this system $M^t_{1\to 2}$ and $M^t_{2\to 1}$ can be calculated analytically. For weak coupling weights it turns out that $e^t_{1\to 2}$ is dominated by $1/|x_1(t)w_{12}|$ and $e^t_{2\to 1}$ by $1/|x_2(t)w_{21}|$. Thus, $M^t_{1\to 2}$ is strongly expansive for low $x_1$ values, and $M^t_{2\to 1}$ for low values of $x_2$. Consequently, as shown in Fig. \ref{fig_sugihara} A) and B), although the coupling weights do not change over time, the asymmetry index $\alpha^t$ fluctuates considerably as the system explores the state space. This change of causal dominance over time gives rise to various dynamical regimes among the time courses of $x_1$ and $x_2$ (Fig. \ref{fig_sugihara} C)). Specifically, it can be seen that when e.g. the influence from $x_1$ to $x_2$ is stronger than in the reverse direction (blue region), i.e. $C^t_{1\to 2} > C^t_{2\to 1}$, the trajectory of $x_1$ is less constrained than the one of $x_2$.

\subsection*{Transitivity, common cause and convergence}
In order to serve as a satisfactory definition of causality in dynamical systems, Topological Causality must meet fundamental requirements that can be demonstrated by examining simple network motifs.

\begin{figure}[h]
	\includegraphics{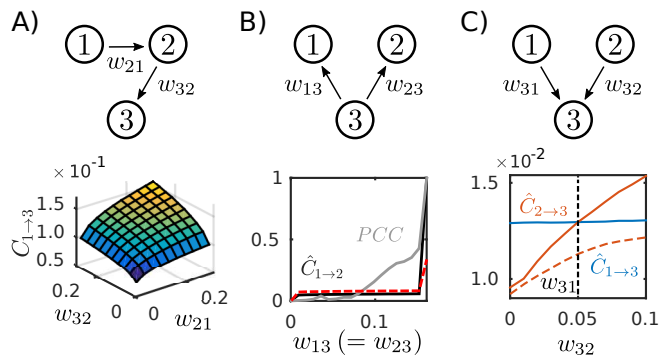}%
	\caption{\label{fig_cause_require} A) Transitivity. A unidirectionally coupled chain $x_1 \to x_2 \to x_3$ is realized by a system of Eqs. (\ref{eq:pawelzik}) $(n=3)$ with only $w_{21} \neq 0$ and $w_{32} \neq 0$. The model was simulated for 10.000 time steps. With an embedding dimension $m=3$, $C^t_{1\to 3}$ can be calculated analytically. One observes $C_{1\to 3} > 0$ if and only if both $w_{21} > 0$ and $w_{32} > 0$. B) Common input. $\hat{C}_{1\to 2}$ is plotted for the same system as in A) but only with $w_{13} = w_{23} \neq 0$. The directed influences $\hat{C}$ are estimated with $k=10$ and an embedding dimension $m=2$. $\hat{C}_{1\to 2}$ and $\hat{C}_{2\to 1}$ (not shown) depend only weakly on the common input and are not significant (the red line is the chance level) unless the Pearson Correlation Coefficient ($PCC$) approaches 1, in which case $x_1$ and $x_2$ become redundant. C) Convergence. The dynamical components 1,2 and 3 are Lorenz systems described by Eqs. (\ref{eq:lorenz}), where only $w_{21}$ and $w_{32}$ are nonzero. For $w_{31} = 0.05$ and varying $w_{32}$, the system with the stronger link to 3 has a higher causal influence $\hat{C}$ (the orange line is the chance level of $\hat{C}_{2\to 3}$). The embedding parameters are $m=16$ and $\tau = 10$, and the local mappings were estimated with $k=2000$ and a full time series length of $10^6$ data points.}
\end{figure}

The first prerequisite is transitivity, meaning that 'if 1 causes 2 and 2 causes 3, then 1 causes 3'. Since $M^t_{3\to 1} = M^t_{2\to 1}M^t_{3\to 2}$, it can be shown that 
\begin{align*}
C^t_{1\to 3} &\geq C^t_{1\to 2}C^t_{2\to 3} \quad \quad \text{if } \quad w_{21}\neq0 \land w_{32}\neq0 \\
C^t_{1\to 3} &= 0  \quad \quad \quad \quad \quad \quad \, \text{else}
\end{align*}
meaning that transitivity is mathematically guaranteed. For a system of 3 coupled logistic maps described by Eqs. (\ref{eq:pawelzik}) $(n = 3)$, and $w_{31} = w_{13} = 0$, the local expansions of the mappings between $x_1$ and $x_3$ can also be calculated analytically. Here, an embedding dimension $m=3$ is sufficient. In the special case of also setting $w_{12} = w_{32} = 0$, resulting in a unidirectional transitive network (Fig. \ref{fig_cause_require} A) top), it turns out that for small couplings $w_{21}$ and $w_{32}$, $e^t_{3\to 1}$ is dominated by $\frac{1}{|w_{21}w_{32}|}$, the product of the small coupling limits of $e^t_{2 \to 1}$ and $e^t_{3 \to 2}$ (compare Eq. (\ref{eq:stretching})). Fig. \ref{fig_cause_require} A) shows the analytical results of $C_{1\to 3}$ for this system for varying coupling weights $w_{21}$ and $w_{32}$.

The second required property is the ability to distinguish shared input from true interaction. Consider a system described by Eqs. (\ref{eq:pawelzik}) $(n=3)$, where only $w_{13} \neq 0$ and $w_{23} \neq 0$, generating a divergent network motif (Fig \ref{fig_cause_require} B) top). With moderate coupling from $x_3$ to $x_1$ and $x_2$, the latter two do not become fully enslaved and, in particular, do not synchronize (which otherwise represents an irrelevant singular case). Fig \ref{fig_cause_require} B) shows estimated values since the theoretical prediction is $C_{1\to 2} = 0$ in any case. The proposed method yields values for the effective influences that are not significant and nearly independent of the common drive, which can induce substantial correlations.

In addition, the measure should be able to deal with convergent influences. Unfortunately, a network of three coupled maps given by Eqs. (\ref{eq:pawelzik}) with a convergent motif such that $x_1$ and $x_2$ are independently influencing $x_3$ cannot be sufficiently embedded. Therefore, convergence needs to be investigated with time continuous component systems for which Takens' theorem is guaranteed to hold. For this purpose three coupled sets of Lorenz equations \citep{Lorenz_1963} are used:
\begin{align}
\label{eq:lorenz}
\dot{x}_{i}(t) &= 10\left(y_{i}(t) - \left(x_{i}(t) - \sum_{j\in\{1,\dots,n\}\backslash \{i\}}w_{ij}x_{j}(t)\right)\right) \\ \nonumber
\dot{y}_{i}(t) &= x_{i}(t)(28 - z_{i}(t)) - y_{i}(t) \\ \nonumber
\dot{z}_{i}(t) &= x_{i}(t)y_{i}(t) - \frac{8}{3}z_{i}(t) \quad \quad ; \quad \quad i=1,\dots,n\\ \nonumber
\end{align}
A convergence motif is achieved when only the coupling weights $w_{31}$ and $w_{32}$ are nonzero. Fig. \ref{fig_cause_require} C) shows that the causal influence of the driving component system with the stronger link to the receiving component system is consistently higher than the influence from the other.

\subsection*{Robustness to observational noise}
\begin{figure}[h!]
	\includegraphics{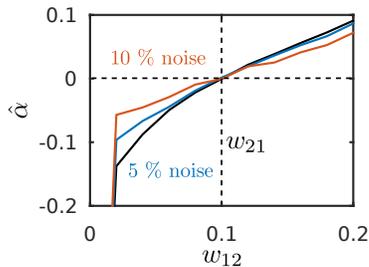}%
	\caption{\label{fig_noise} The values of the mean asymmetry index $\hat{\alpha}$ are shown for two coupled Lorenz-systems described by Eqs. (\ref{eq:lorenz}) ($n = 2$) with a fixed coupling $w_{21} = 0.1$ and a range of different values for $w_{12}$. 150.000 data points were used with $m=13$ and $k=200$. The black line displays the result for the noiseless case. The two estimates under conditions of additive Gaussian white noise (5\% and 10\%) still succeed to capture the asymmetry of influences quantitatively except for low values of $w_{12}$. We set non-significant $\hat{C}$ to 0 leading to $\hat \alpha = -1$ for small $w_{12}$.}
\end{figure}

To demonstrate the robustness to noise, the dependency of the asymmetry index $\alpha$ on additive Gaussian noise is investigated for two coupled Lorenz systems described by Eqs. (\ref{eq:lorenz}) $(n=2)$. Fig. \ref{fig_noise} shows $\hat{\alpha}$ for various combinations of $w_{21}$ and $w_{12}$ in a noise-free case and for 5\% and 10\% of noise. While the direction of dominant influence is faithfully reproduced, the manifold structure in the chosen neighborhood size is partially masked by the noise. 

\subsection*{Linear systems with intrinsic noise}
\begin{figure}[t]
	\includegraphics{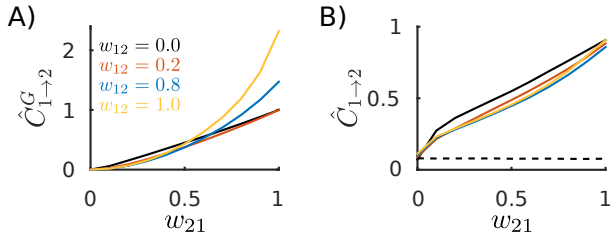}
	\caption{\label{fig_granger_vs_topo} Numerical results are shown for Wiener-Granger-Causality (A)) and Topological Causality (B)) based on the same time series generated by Eqs. \ref{eq:granger} with $\beta=0.95$ for different combinations of coupling weights $w_{12}$ and $w_{21}$. For the estimation of each value $10^6$ data points were used. The dashed line in B) is the chance level.}
\end{figure}

Noise can hamper detectability of local topological structure. However, also global structure can convey information about causal relationships. As a prominent example consider a two dimensional linear system driven by noise:
\begin{align}
\label{eq:granger}
x_1(t+1) &= \beta[(1-w_{12})x_1(t) + w_{12}x_2(t)] + \eta_{1}(t) \\ \nonumber
x_2(t+1) &= \beta[(1-w_{21})x_2(t) + w_{21}x_1(t)] + \eta_{2}(t) \\ \nonumber
\end{align}
with $\beta < 1$ and $\eta_{1}$ and $\eta_{2}$ denoting uncorrelated Gaussian white noise processes. Since interaction between the two component systems is linear, average mappings between reconstruction spaces will also be linear and consequently not state dependent. Thus, to apply Topological Causality to such a system, global mappings $M_{1\to 2}$ and $M_{2\to 1}$ can be fitted using the full ensemble of available data points. Since this is a classical case for which Wiener-Granger Causality \citep{Wiener_1956,Granger_1969} is designed, Fig. \ref{fig_granger_vs_topo} A) shows the estimated Wiener-Granger Causality $\hat{C}^{G}_{1\to 2}$ using one previous time step for prediction for different combinations of coupling weights. To obtain comparable results, $\hat{C}_{1\to 2}$ is estimated from the mapping that projects $r^{x_2}(t+1)$ to $r^{x_1}(t)$ with $m = 2$ (Fig. \ref{fig_granger_vs_topo} B). While $\hat{C}^{G}_{1\to 2}$ depends substantially also on $w_{12}$, $\hat{C}_{1\to 2}$ more faithfully reflects the true underlying interaction strength $w_{21}$ for the full range of both parameters.

\section*{Discussion}
The definition of Topological Causality (TC) we put forward here is tailored to the detection of directed effective influences among mutually coupled dynamical systems. It follows heuristic considerations of influence-induced distortions in the mappings between manifold reconstructions and is linked to entropy production. It relies on the expansion of the mappings, which is in contrast to methods that use the full log determinant \citep{Schoelkopf_2012}. In simple cases the causality index $C^t$ is found to be fully analytically tractable and shown to reflect directed effective influences of system components including their state and time dependence. In more complex systems (as e.g. Eq. 5) $C$ reflects the compound influence of one observable onto another exerted along multiple and possibly cyclic paths through the network of coupled components. Importantly, TC is demonstrated to fulfill basic requirements that a measure of causality must obey, such as transitivity and disentanglement of causal influence from common input. 

To overcome limitations of Wiener-Granger Causality (WGC) \citep{Wiener_1956,Granger_1969} several approaches for evaluating causal interactions in non-separable dynamical systems have also been based on relations among state-space reconstructions. For example, tests for the existence of directed unique mappings between reconstructed manifolds can be used as an all or nothing criterion to detect causal links between component systems \citep{Pawelzik_1991, Chicarro_2009, Ma_2014, Sugihara_2012, Gedeon_2015}. 

The current method for estimating TC from observed time series is most closely related to the empirical procedure of Convergent Cross-Mapping (CCM) \citep{Sugihara_2012} that yield interesting results in a range of applications, e.g. \citep{Sugihara_2012,Wang_2014,Tajima_2015,Nes_2015}. The proposed gradual measure of the causal influence relies on the errors when predicting one reconstruction manifold from another: the slower the convergence of the prediction error of $r^{x_1}$ from $r^{x_2}$ with increasing time series length, the weaker the causation $x_1$ to $x_2$. We suggest that this effect is a consequence of the expansion which TC measures directly: the more expansive the mapping $M_{2 \to 1}$ locally is, the more its non-linearities will hamper predictions with a given finite number of data points. In other words, we believe CCM evaluates deviations from the assumption that the mapping $\{r^{x_2}(t^{x_{2}}_{1}), ..., r^{x_2}(t^{x_{2}}_{k})\}$ to $\{r^{x_1}(t^{x_{2}}_{1}), ..., r^{x_1}(t^{x_{2}}_{k})\}$
is linear and therefore is an indirect measure of the underlying directed effective influence. 

Being a concept tailored to non-separable deterministic systems which preserve information between coupled components, TC seems complementary to methods for determining causal influences in stochastic systems. Most prominent examples are WGC and Transfer Entropy (TE) \citep{Schreiber_2000}, which are conceptually related \citep{Barnett_2009}. Both are based on the reduction of uncertainty in one time series by including past information from the other. It might therefore come as a surprise that TC can detect effective influences also in predominantly stochastic linear systems (Fig. \ref{fig_granger_vs_topo}). However, both approaches are not independent: In stochastic linear systems, the observed dynamics in the reconstruction spaces can be interpreted as samples from a probability density. Heuristically, an expansive mapping between probability densities increases entropy and thereby induces information loss. In stark contrast to the usual applications of WGC and TE, however, TC exploits the expansion of the {\em backward} mapping from 'effect' $x_2$ to 'cause' $x_1$ for determining the causal influence from $x_1$ to $x_2$. In terms of uncertainty reduction this would correspond to stating: 'If state reconstructions of $x_1$ can be better determined by taking future state reconstructions of $x_2$ into account, a causal link from $x_1$ to $x_2$ exists'. Since this is possible in TC also a time reversed application of WGC should reflect the causal link of $x_1$ on $x_2$: Influence from $x_1$ to $x_2$ transports information about $x_1$ to later states of $x_2$ that could be used to 'postdict' previous states of $x_1$. This intuition was in fact already applied \citep{Haufe_2012}. This raises the intriguing possibility that TC could be well suited for both deterministic and stochastic systems, where in the first case it exploits characteristics of entangled systems, and in the latter approaches results of WGC and TE.

For practical applications of TC, care has to be taken with respect to choice of neighborhood, embedding parameters and fit procedures. Given large numbers of noiseless data points, the method used in this paper only requires to fit linear mappings. For finite time series contaminated with noise, however, useful neighborhood scales will depend on noise levels and available time series length. Also systematic biases due to the manifolds’ geometry and density as well as noise induced biases will need to be accounted for. The details of practical procedures that achieve significant results for real data remain to be explored which is beyond the scope of this paper.

\begin{acknowledgments}
DH was funded by the Bundesministerium f\"ur Bildung und Forschung (Bernstein Award Udo Ernst, Grant 01GQ1106) and thanks UA Ernst for support during the project. We thank D Rotermund, UA Ernst and M Sch\"unemann for helpful comments on the manuscript.
\end{acknowledgments}

%---\newpage
\bibliography{topo_cause_refs.bib}

\begin{thebibliography}{20}%
\makeatletter
\providecommand \@ifxundefined [1]{%
 \@ifx{#1\undefined}
}%
\providecommand \@ifnum [1]{%
 \ifnum #1\expandafter \@firstoftwo
 \else \expandafter \@secondoftwo
 \fi
}%
\providecommand \@ifx [1]{%
 \ifx #1\expandafter \@firstoftwo
 \else \expandafter \@secondoftwo
 \fi
}%
\providecommand \natexlab [1]{#1}%
\providecommand \enquote  [1]{``#1''}%
\providecommand \bibnamefont  [1]{#1}%
\providecommand \bibfnamefont [1]{#1}%
\providecommand \citenamefont [1]{#1}%
\providecommand \href@noop [0]{\@secondoftwo}%
\providecommand \href [0]{\begingroup \@sanitize@url \@href}%
\providecommand \@href[1]{\@@startlink{#1}\@@href}%
\providecommand \@@href[1]{\endgroup#1\@@endlink}%
\providecommand \@sanitize@url [0]{\catcode `\\12\catcode `\$12\catcode
  `\&12\catcode `\#12\catcode `\^12\catcode `\_12\catcode `\%12\relax}%
\providecommand \@@startlink[1]{}%
\providecommand \@@endlink[0]{}%
\providecommand \url  [0]{\begingroup\@sanitize@url \@url }%
\providecommand \@url [1]{\endgroup\@href {#1}{\urlprefix }}%
\providecommand \urlprefix  [0]{URL }%
\providecommand \Eprint [0]{\href }%
\providecommand \doibase [0]{http://dx.doi.org/}%
\providecommand \selectlanguage [0]{\@gobble}%
\providecommand \bibinfo  [0]{\@secondoftwo}%
\providecommand \bibfield  [0]{\@secondoftwo}%
\providecommand \translation [1]{[#1]}%
\providecommand \BibitemOpen [0]{}%
\providecommand \bibitemStop [0]{}%
\providecommand \bibitemNoStop [0]{.\EOS\space}%
\providecommand \EOS [0]{\spacefactor3000\relax}%
\providecommand \BibitemShut  [1]{\csname bibitem#1\endcsname}%
\let\auto@bib@innerbib\@empty
%</preamble>
\bibitem [{\citenamefont {Aristotle}(0 BC)}]{aristotle}%
  \BibitemOpen
  \bibfield  {author} {\bibinfo {author} {\bibnamefont {Aristotle}},\
  }\href@noop {} {\bibfield  {journal} {\bibinfo  {journal} {Metaphysics}\ }
  (\bibinfo {year} {350 BC})}\BibitemShut {NoStop}%
\bibitem [{\citenamefont {Ay}\ and\ \citenamefont {Polani}(2008)}]{Ay_2008}%
  \BibitemOpen
  \bibfield  {author} {\bibinfo {author} {\bibfnamefont {N.}~\bibnamefont
  {Ay}}\ and\ \bibinfo {author} {\bibfnamefont {D.}~\bibnamefont {Polani}},\
  }\href@noop {} {\bibfield  {journal} {\bibinfo  {journal} {Advances in
  Complex Systems}\ }\textbf {\bibinfo {volume} {11}},\ \bibinfo {pages} {17}
  (\bibinfo {year} {2008})}\BibitemShut {NoStop}%
\bibitem [{\citenamefont {Takens}(1981)}]{Takens_1981}%
  \BibitemOpen
  \bibfield  {author} {\bibinfo {author} {\bibfnamefont {F.}~\bibnamefont
  {Takens}},\ }in\ \href@noop {} {\emph {\bibinfo {booktitle} {Dynamical
  Systems and Turbulence}}},\ \bibinfo {series} {Springer Lecture Notes in
  Mathematics}, Vol.\ \bibinfo {volume} {898},\ \bibinfo {editor} {edited by\
  \bibinfo {editor} {\bibfnamefont {D.~A.}\ \bibnamefont {Rand}}\ and\ \bibinfo
  {editor} {\bibfnamefont {L.-S.}\ \bibnamefont {Young}}}\ (\bibinfo
  {publisher} {Springer-Verlag},\ \bibinfo {address} {Berlin},\ \bibinfo {year}
  {1981})\BibitemShut {NoStop}%
\bibitem [{\citenamefont {Packard}\ \emph {et~al.}(1980)\citenamefont
  {Packard}, \citenamefont {Crutchfield}, \citenamefont {Farmer},\ and\
  \citenamefont {Shaw}}]{Packard_1980}%
  \BibitemOpen
  \bibfield  {author} {\bibinfo {author} {\bibfnamefont {N.~H.}\ \bibnamefont
  {Packard}}, \bibinfo {author} {\bibfnamefont {J.~P.}\ \bibnamefont
  {Crutchfield}}, \bibinfo {author} {\bibfnamefont {J.~D.}\ \bibnamefont
  {Farmer}}, \ and\ \bibinfo {author} {\bibfnamefont {R.~S.}\ \bibnamefont
  {Shaw}},\ }\href@noop {} {\bibfield  {journal} {\bibinfo  {journal} {Phys.
  Rev. Lett.}\ }\textbf {\bibinfo {volume} {45}},\ \bibinfo {pages} {712}
  (\bibinfo {year} {1980})}\BibitemShut {NoStop}%
\bibitem [{\citenamefont {Anderson}(1984)}]{Anderson_1984}%
  \BibitemOpen
  \bibfield  {author} {\bibinfo {author} {\bibfnamefont {T.~W.}\ \bibnamefont
  {Anderson}},\ }\href@noop {} {\emph {\bibinfo {title} {An Introduction to
  Multivariate Statistical Analysis}}},\ \bibinfo {edition} {2nd}\ ed.\
  (\bibinfo  {publisher} {Wiley},\ \bibinfo {address} {New York},\ \bibinfo
  {year} {1984})\BibitemShut {NoStop}%
\bibitem [{\citenamefont {Sugihara}\ \emph {et~al.}(2012)\citenamefont
  {Sugihara}, \citenamefont {May}, \citenamefont {Ye}, \citenamefont {Hsieh},
  \citenamefont {Deyle},\ and\ \citenamefont {Fogarty}}]{Sugihara_2012}%
  \BibitemOpen
  \bibfield  {author} {\bibinfo {author} {\bibfnamefont {G.}~\bibnamefont
  {Sugihara}}, \bibinfo {author} {\bibfnamefont {R.~M.}\ \bibnamefont {May}},
  \bibinfo {author} {\bibfnamefont {H.}~\bibnamefont {Ye}}, \bibinfo {author}
  {\bibfnamefont {C.}~\bibnamefont {Hsieh}}, \bibinfo {author} {\bibfnamefont
  {E.~R.}\ \bibnamefont {Deyle}}, \ and\ \bibinfo {author} {\bibfnamefont
  {M.}~\bibnamefont {Fogarty}},\ }\href@noop {} {\bibfield  {journal} {\bibinfo
   {journal} {Science}\ }\textbf {\bibinfo {volume} {334}},\ \bibinfo {pages}
  {496} (\bibinfo {year} {2012})}\BibitemShut {NoStop}%
\bibitem [{\citenamefont {Lorenz}(1963)}]{Lorenz_1963}%
  \BibitemOpen
  \bibfield  {author} {\bibinfo {author} {\bibfnamefont {E.~N.}\ \bibnamefont
  {Lorenz}},\ }\href@noop {} {\bibfield  {journal} {\bibinfo  {journal}
  {Journal of the Atmospheric Sciences}\ }\textbf {\bibinfo {volume} {20}},\
  \bibinfo {pages} {130} (\bibinfo {year} {1963})}\BibitemShut {NoStop}%
\bibitem [{\citenamefont {Wiener}(1956)}]{Wiener_1956}%
  \BibitemOpen
  \bibfield  {author} {\bibinfo {author} {\bibfnamefont {N.}~\bibnamefont
  {Wiener}},\ }in\ \href@noop {} {\emph {\bibinfo {booktitle} {Modern
  Mathematics for Engineers}}},\ \bibinfo {editor} {edited by\ \bibinfo
  {editor} {\bibfnamefont {E.~F.}\ \bibnamefont {Beckenbach}}}\ (\bibinfo
  {publisher} {McGraw-Hill},\ \bibinfo {address} {New York},\ \bibinfo {year}
  {1956})\BibitemShut {NoStop}%
\bibitem [{\citenamefont {Granger}(1969)}]{Granger_1969}%
  \BibitemOpen
  \bibfield  {author} {\bibinfo {author} {\bibfnamefont {C.~W.~J.}\
  \bibnamefont {Granger}},\ }\href@noop {} {\bibfield  {journal} {\bibinfo
  {journal} {Econometrica}\ }\textbf {\bibinfo {volume} {37}},\ \bibinfo
  {pages} {424} (\bibinfo {year} {1969})}\BibitemShut {NoStop}%
\bibitem [{\citenamefont {Janzing}\ \emph {et~al.}(2012)\citenamefont
  {Janzing}, \citenamefont {Mooij}, \citenamefont {Zhang}, \citenamefont
  {Lemeire}, \citenamefont {Zscheischler}, \citenamefont {Daniu{\v{s}}sis},
  \citenamefont {Steudel},\ and\ \citenamefont
  {Sch{\"{o}}lkopf}}]{Schoelkopf_2012}%
  \BibitemOpen
  \bibfield  {author} {\bibinfo {author} {\bibfnamefont {D.}~\bibnamefont
  {Janzing}}, \bibinfo {author} {\bibfnamefont {J.}~\bibnamefont {Mooij}},
  \bibinfo {author} {\bibfnamefont {K.}~\bibnamefont {Zhang}}, \bibinfo
  {author} {\bibfnamefont {J.}~\bibnamefont {Lemeire}}, \bibinfo {author}
  {\bibfnamefont {J.}~\bibnamefont {Zscheischler}}, \bibinfo {author}
  {\bibfnamefont {P.}~\bibnamefont {Daniu{\v{s}}sis}}, \bibinfo {author}
  {\bibfnamefont {B.}~\bibnamefont {Steudel}}, \ and\ \bibinfo {author}
  {\bibfnamefont {B.}~\bibnamefont {Sch{\"{o}}lkopf}},\ }\href@noop {}
  {\bibfield  {journal} {\bibinfo  {journal} {Artificial Intelligence}\
  }\textbf {\bibinfo {volume} {182--183}},\ \bibinfo {pages} {1} (\bibinfo
  {year} {2012})}\BibitemShut {NoStop}%
\bibitem [{\citenamefont {Liebert}\ \emph {et~al.}(1991)\citenamefont
  {Liebert}, \citenamefont {Pawelzik},\ and\ \citenamefont
  {Schuster}}]{Pawelzik_1991}%
  \BibitemOpen
  \bibfield  {author} {\bibinfo {author} {\bibfnamefont {W.}~\bibnamefont
  {Liebert}}, \bibinfo {author} {\bibfnamefont {K.~R.}\ \bibnamefont
  {Pawelzik}}, \ and\ \bibinfo {author} {\bibfnamefont {H.~G.}\ \bibnamefont
  {Schuster}},\ }\href@noop {} {\bibfield  {journal} {\bibinfo  {journal}
  {Europhysics Letters}\ }\textbf {\bibinfo {volume} {14}},\ \bibinfo {pages}
  {521} (\bibinfo {year} {1991})}\BibitemShut {NoStop}%
\bibitem [{\citenamefont {Chicharro}\ and\ \citenamefont
  {Andrzejak}(2009)}]{Chicarro_2009}%
  \BibitemOpen
  \bibfield  {author} {\bibinfo {author} {\bibfnamefont {D.}~\bibnamefont
  {Chicharro}}\ and\ \bibinfo {author} {\bibfnamefont {R.~G.}\ \bibnamefont
  {Andrzejak}},\ }\href@noop {} {\bibfield  {journal} {\bibinfo  {journal}
  {Phys. Rev. E}\ }\textbf {\bibinfo {volume} {80}},\ \bibinfo {pages} {026217}
  (\bibinfo {year} {2009})}\BibitemShut {NoStop}%
\bibitem [{\citenamefont {Ma}\ \emph {et~al.}(2014)\citenamefont {Ma},
  \citenamefont {Aihara},\ and\ \citenamefont {Chen}}]{Ma_2014}%
  \BibitemOpen
  \bibfield  {author} {\bibinfo {author} {\bibfnamefont {H.}~\bibnamefont
  {Ma}}, \bibinfo {author} {\bibfnamefont {K.}~\bibnamefont {Aihara}}, \ and\
  \bibinfo {author} {\bibfnamefont {L.}~\bibnamefont {Chen}},\ }\href@noop {}
  {\bibfield  {journal} {\bibinfo  {journal} {Scientific Reports}\ }\textbf
  {\bibinfo {volume} {4}},\ \bibinfo {pages} {7464} (\bibinfo {year}
  {2014})}\BibitemShut {NoStop}%
\bibitem [{\citenamefont {Cummins}\ \emph {et~al.}(2015)\citenamefont
  {Cummins}, \citenamefont {Gedeon},\ and\ \citenamefont
  {Spendlove}}]{Gedeon_2015}%
  \BibitemOpen
  \bibfield  {author} {\bibinfo {author} {\bibfnamefont {B.}~\bibnamefont
  {Cummins}}, \bibinfo {author} {\bibfnamefont {T.}~\bibnamefont {Gedeon}}, \
  and\ \bibinfo {author} {\bibfnamefont {K.}~\bibnamefont {Spendlove}},\
  }\href@noop {} {\bibfield  {journal} {\bibinfo  {journal} {SIAM J. Applied
  Dynamical Systems}\ }\textbf {\bibinfo {volume} {14}},\ \bibinfo {pages}
  {335} (\bibinfo {year} {2015})}\BibitemShut {NoStop}%
\bibitem [{\citenamefont {Wang}\ \emph {et~al.}(2014)\citenamefont {Wang},
  \citenamefont {Piao}, \citenamefont {Ciais}, \citenamefont {Friedlingstein},
  \citenamefont {Myneni}, \citenamefont {Cox}, \citenamefont {Heimann},
  \citenamefont {Miller}, \citenamefont {Peng}, \citenamefont {Wang},
  \citenamefont {Yanga},\ and\ \citenamefont {Chen}}]{Wang_2014}%
  \BibitemOpen
  \bibfield  {author} {\bibinfo {author} {\bibfnamefont {X.}~\bibnamefont
  {Wang}}, \bibinfo {author} {\bibfnamefont {S.}~\bibnamefont {Piao}}, \bibinfo
  {author} {\bibfnamefont {P.}~\bibnamefont {Ciais}}, \bibinfo {author}
  {\bibfnamefont {P.}~\bibnamefont {Friedlingstein}}, \bibinfo {author}
  {\bibfnamefont {R.~B.}\ \bibnamefont {Myneni}}, \bibinfo {author}
  {\bibfnamefont {P.}~\bibnamefont {Cox}}, \bibinfo {author} {\bibfnamefont
  {M.}~\bibnamefont {Heimann}}, \bibinfo {author} {\bibfnamefont
  {J.}~\bibnamefont {Miller}}, \bibinfo {author} {\bibfnamefont
  {S.}~\bibnamefont {Peng}}, \bibinfo {author} {\bibfnamefont {T.}~\bibnamefont
  {Wang}}, \bibinfo {author} {\bibfnamefont {H.}~\bibnamefont {Yanga}}, \ and\
  \bibinfo {author} {\bibfnamefont {A.}~\bibnamefont {Chen}},\ }\href@noop {}
  {\bibfield  {journal} {\bibinfo  {journal} {Nature}\ }\textbf {\bibinfo
  {volume} {506}},\ \bibinfo {pages} {212} (\bibinfo {year}
  {2014})}\BibitemShut {NoStop}%
\bibitem [{\citenamefont {Tajima}\ \emph {et~al.}(2015)\citenamefont {Tajima},
  \citenamefont {Yanagawa}, \citenamefont {Fujii},\ and\ \citenamefont
  {Toyoizumi}}]{Tajima_2015}%
  \BibitemOpen
  \bibfield  {author} {\bibinfo {author} {\bibfnamefont {S.}~\bibnamefont
  {Tajima}}, \bibinfo {author} {\bibfnamefont {T.}~\bibnamefont {Yanagawa}},
  \bibinfo {author} {\bibfnamefont {N.}~\bibnamefont {Fujii}}, \ and\ \bibinfo
  {author} {\bibfnamefont {T.}~\bibnamefont {Toyoizumi}},\ }\href@noop {}
  {\bibfield  {journal} {\bibinfo  {journal} {PLoS Comp. Biol.}\ }\textbf
  {\bibinfo {volume} {1004537}} (\bibinfo {year} {2015})}\BibitemShut {NoStop}%
\bibitem [{\citenamefont {van Nes}\ \emph {et~al.}(2015)\citenamefont {van
  Nes}, \citenamefont {Scheffer}, \citenamefont {Brovkin}, \citenamefont
  {Lenton}, \citenamefont {Ye}, \citenamefont {Deyle},\ and\ \citenamefont
  {Sugihara}}]{Nes_2015}%
  \BibitemOpen
  \bibfield  {author} {\bibinfo {author} {\bibfnamefont {E.~H.}\ \bibnamefont
  {van Nes}}, \bibinfo {author} {\bibfnamefont {M.}~\bibnamefont {Scheffer}},
  \bibinfo {author} {\bibfnamefont {V.}~\bibnamefont {Brovkin}}, \bibinfo
  {author} {\bibfnamefont {T.~M.}\ \bibnamefont {Lenton}}, \bibinfo {author}
  {\bibfnamefont {H.}~\bibnamefont {Ye}}, \bibinfo {author} {\bibfnamefont
  {E.}~\bibnamefont {Deyle}}, \ and\ \bibinfo {author} {\bibfnamefont
  {G.}~\bibnamefont {Sugihara}},\ }\href@noop {} {\bibfield  {journal}
  {\bibinfo  {journal} {Nature climate change}\ }\textbf {\bibinfo {volume}
  {5}},\ \bibinfo {pages} {DOI: 10.1038/NCLIMATE2568} (\bibinfo {year}
  {2015})}\BibitemShut {NoStop}%
\bibitem [{\citenamefont {Schreiber}(2000)}]{Schreiber_2000}%
  \BibitemOpen
  \bibfield  {author} {\bibinfo {author} {\bibfnamefont {T.}~\bibnamefont
  {Schreiber}},\ }\href@noop {} {\bibfield  {journal} {\bibinfo  {journal}
  {Phys. Rev. Lett.}\ }\textbf {\bibinfo {volume} {85}},\ \bibinfo {pages}
  {461} (\bibinfo {year} {2000})}\BibitemShut {NoStop}%
\bibitem [{\citenamefont {Barnett}\ \emph {et~al.}(2009)\citenamefont
  {Barnett}, \citenamefont {Barrett},\ and\ \citenamefont
  {Seth}}]{Barnett_2009}%
  \BibitemOpen
  \bibfield  {author} {\bibinfo {author} {\bibfnamefont {L.}~\bibnamefont
  {Barnett}}, \bibinfo {author} {\bibfnamefont {A.~B.}\ \bibnamefont
  {Barrett}}, \ and\ \bibinfo {author} {\bibfnamefont {A.~K.}\ \bibnamefont
  {Seth}},\ }\href@noop {} {\bibfield  {journal} {\bibinfo  {journal} {Phys.
  Rev. Lett.}\ }\textbf {\bibinfo {volume} {103}},\ \bibinfo {pages} {238701}
  (\bibinfo {year} {2009})}\BibitemShut {NoStop}%
\bibitem [{\citenamefont {Haufe}\ \emph {et~al.}(2012)\citenamefont {Haufe},
  \citenamefont {Nikulin}, \citenamefont {M{\"u}ller},\ and\ \citenamefont
  {Nolte}}]{Haufe_2012}%
  \BibitemOpen
  \bibfield  {author} {\bibinfo {author} {\bibfnamefont {S.}~\bibnamefont
  {Haufe}}, \bibinfo {author} {\bibfnamefont {V.~V.}\ \bibnamefont {Nikulin}},
  \bibinfo {author} {\bibfnamefont {K.-R.}\ \bibnamefont {M{\"u}ller}}, \ and\
  \bibinfo {author} {\bibfnamefont {G.}~\bibnamefont {Nolte}},\ }\href@noop {}
  {\bibfield  {journal} {\bibinfo  {journal} {Neuroimage}\ }\textbf {\bibinfo
  {volume} {64}},\ \bibinfo {pages} {120} (\bibinfo {year} {2012})}\BibitemShut
  {NoStop}%
\end{thebibliography}%

\end{document}